\documentclass[12pt]{article}
\usepackage{graphicx}
\usepackage{color}

\begin{document}
\title{On the possibility to use semiconductive hybrid pixel detectors for study of radiation belt  of the Earth.\\
{\small ICPPA-2015 proceedings}}
\author{A. Guskov$^{1,*}$, G. Shelkov$^{1}$, P. Smolyanskiy$^{1}$, A. Zhemchugov$^{1}$}
\maketitle
$^1$Joint Institute for Nuclear Research, Dubna

$^*$avg@jinr.ru

\begin{abstract}
The scientific apparatus ''Gamma-400'' designed for study of hadron and electromagnetic components of cosmic rays will be launched to an elliptic orbit with the apogee of about 300 000 km and the perigee of about 500 km. Such a configuration of the orbit allows it to cross periodically the radiation belt and the outer part of magnetosphere. We discuss the possibility to use hybrid pixel detecters based on the Timepix chip and semiconductive sensors on board the ''Gamma-400'' apparatus. Due to high granularity of the sensor (pixel size is 55 $mu$m) and possibility to measure independently an energy deposition in each pixel, such compact and lightweight detector could be a unique instrument for study of spatial, energy and time structure of electron and proton components of the radiation belt.
\end{abstract}
\maketitle

\section{Timepix detectors}
Pixel hybrid detectors consisted of the Timepix chip \cite{timepix1}, developed at CERN, coupled with a semiconductive sensor are used as a particle detector in a variety of disciplines from the study of cosmic rays to biomedical imaging. Good spatial (pixel size is 55 $\mu m$) and energy (energy deposition can be measured individually in each pixel)  resolution, low level of noise, compact size and radiation hardness are doubtless advantages of this type of detectors. Thin plates (0.3-1.0 mm) of 
silicon (Si) or gallium arsenide compensated with chromium (GaAs) with 256$\times$256 square pixels are used as a sensor. GaAs sensors are developed and produced in the Tomsk State University (Tomsk, Russia)\cite{GaAs1,GaAs2,GaAs3,GaAs4}. The Joint Institute for Nuclear Research (Dubna, Russia) has wide experience in the production and operation of  such kind of the detectors since 2008 \cite{JINR1,JINR2}. 

\begin{figure}[h]
\begin{minipage}{18pc}
\includegraphics[width=18pc]{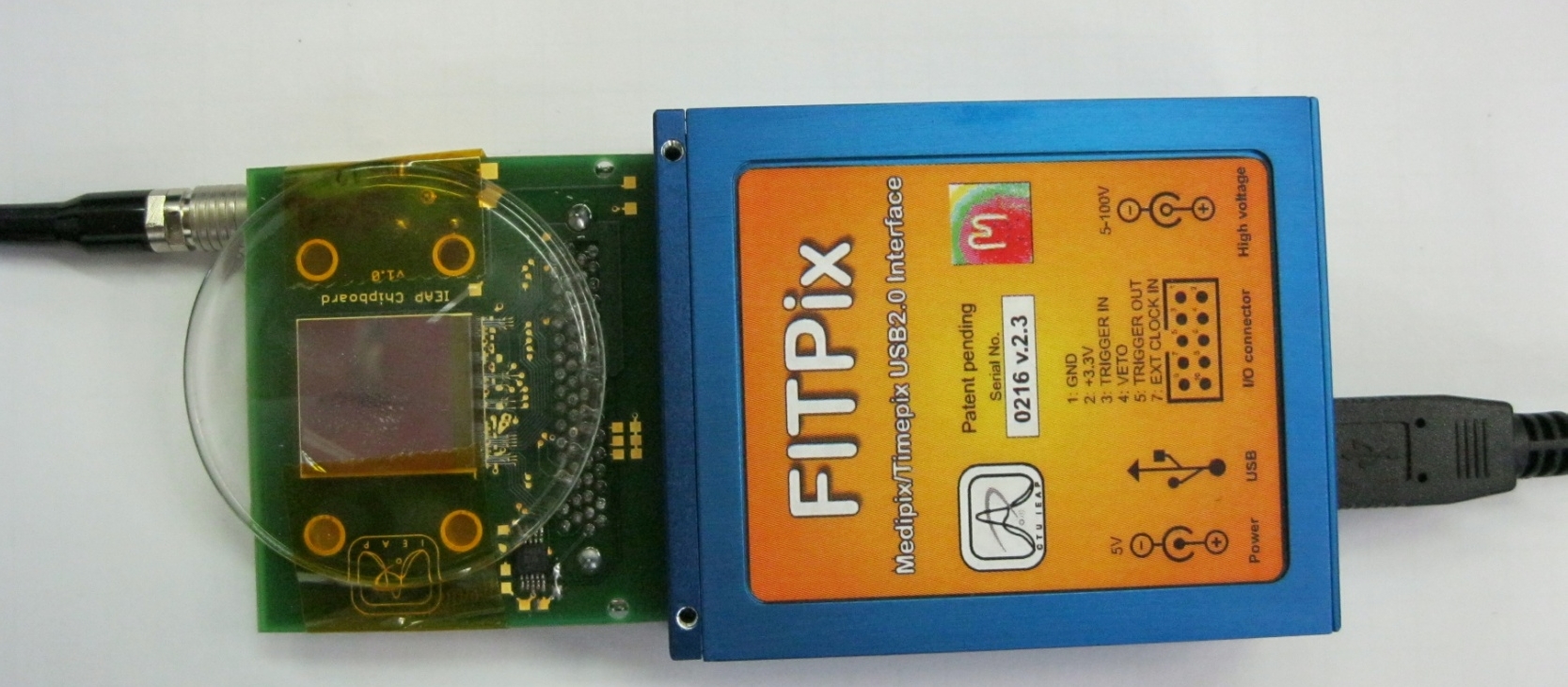}
\caption{\label{label1}Layout of the Timepix detector}
\end{minipage}\hspace{2pc}%
\begin{minipage}{18pc}
\includegraphics[width=18pc]{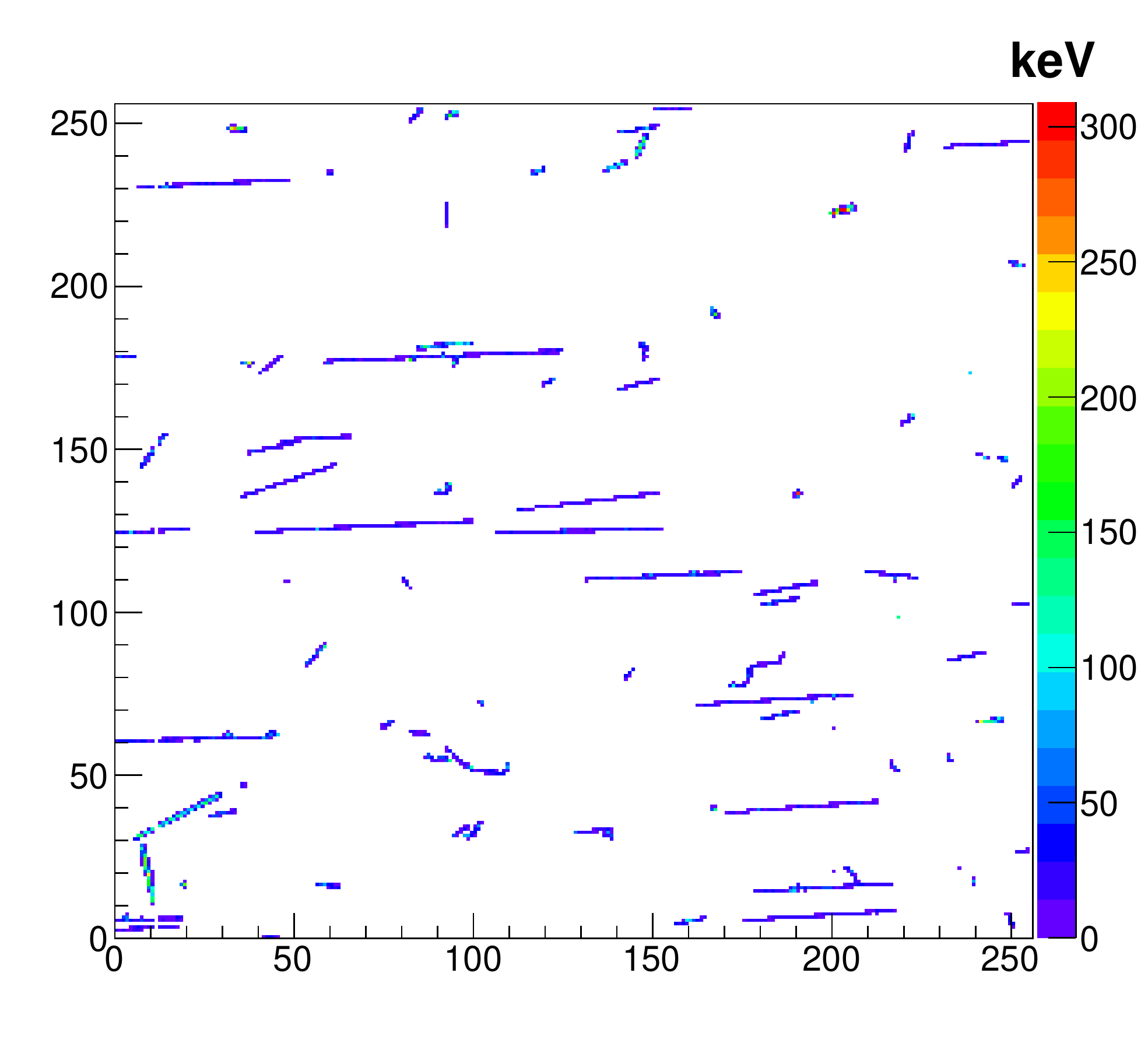}
\caption{\label{label2}Response of the Timepix detector equipped with 0.3 mm silicon sensor placed in the vicinity of the interaction point of 2 GeV deutron beam with lead target.}
\end{minipage} 
\end{figure}

\begin{figure}[h]
\begin{minipage}{18pc}
\includegraphics[width=18pc]{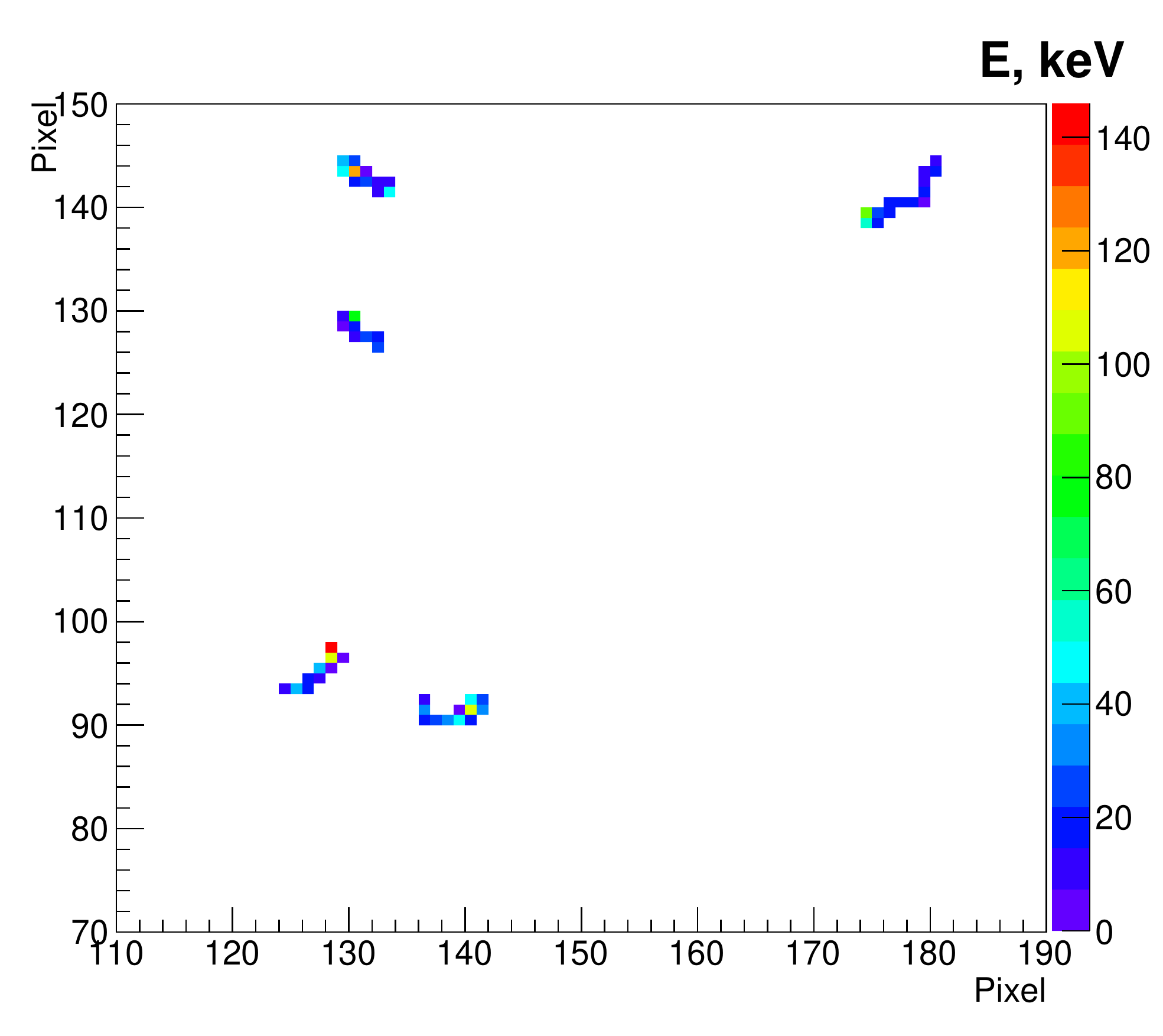}
\end{minipage}\hspace{2pc}%
\begin{minipage}{18pc}
\includegraphics[width=18pc]{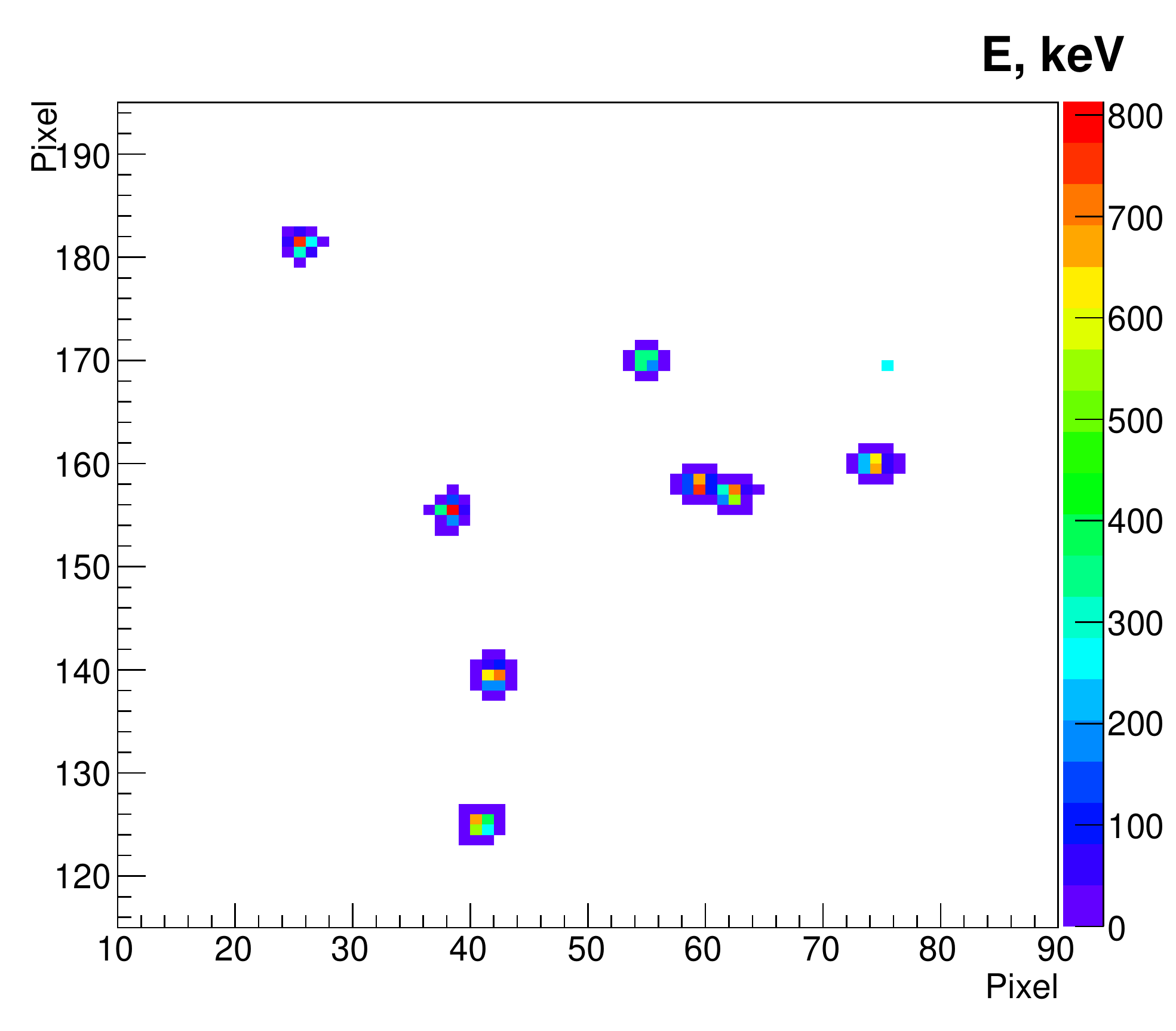}
\end{minipage} 
\caption{\label{label3}Typical responses of the Timepix detector equipped with 0.3 mm silicon sensor and irradiated with 350 keV electrons (left) and with 2 MeV protons (right). Angle of incidence is $90^{o}$. Only a part of the frame is shown.}
\end{figure}

Layout of the Timepix detector is presented in Fig. \ref{label1}. Sensitivity of the detector starts from about 6 keV for $\gamma$-quants, from 30 keV for electrons and from 500 keV for protons. Interaction of an incoming particle with the material of the sensor produces a cluster of energy deposition. Analysis of geometrical shape of such cluster and energy distribution over pixels provides possibility to classify clusters as produced by soft photons, low-energetic electrons, non-relativistic protons, neutrons, heavy ions and minimum ionizing particles (ultrarelativistic charged particles). Fig \ref{label2}. shows response of the Timepix detector with 0.3 mm silicon sensor placed in the vicinity of the interaction point of 2 GeV deutron beam with lead target. Long straight horizontal lines correspond to the beam halo deutrons.Typical responses of the Timepix detector, equipped with 0.3 mm silicon sensor, irradiated with 350 keV electrons and with 2 MeV protons, coming to the surface at $90^{o}$ angle, are presented on Fig. \ref{label3} (left) and (right) correspondently. Since for lowenergetic particles full energy deposition occurs inside the sensor,  the energy of incoming electrons and protons can be measured at least up to 0.3 MeV and 6 MeV in 0.3 mm silicon and up to 1.4 MeV and 17 MeV in 1 mm gallium arsenide sensors correspondently.

Basic properties of the Timepix detectors are presented in Tab. \ref{parameters}. 

\begin{table}[b]
\centering
\caption{\label{parameters}Basic parameters of the Timepix detector.}
\begin{tabular}{@{}l*{15}{l}}
\hline
Parameter & Value\\
\hline
Mass, g & 200\\
Dimensions $L\times W\times H$, cm &  $15\times 5\times 3$\\
Power consumption, W & 2.5\\
Radiation hardness  & \\
of Timepix chip, MGy &4.6 \cite{RAD_Timepix}\\
Radiation hardness  & \\
of Si sensor, MGy & 0.1-0.5 \\
Radiation hardness  & \\
of GaAs sensor, MGy & 1.5 \cite{RADGAAS}\\
Electric field strength, V/$\mu$m &  0.5-1.0 \\
Minimal frame length, ms & 0.01 \\
Amount of data, kB per frame & \\
with occupancy  10\% &  20 \\
Maximal frame rate, kHz & 100 \\
Sensitivity range, keV & $>$6($\gamma$),$>$30(e),$>$500(p)\\
Working area, cm$^2$ & 2 \\
Interface & USB \\
\hline
\end{tabular}
\end{table}

\section{Timepix detectors at ''Gamma-400''}
 There is already a positive experience to use the Timepix detectors in space. It is used on board the International Space Station to accurately monitor radiation doses from various sources \cite{ISS}.  NASA has used the Timepix detector during the first test flight of the ''Orion module'' - part of the next generation of manned US spacecraft \cite{Orion}. It was also installed on board the ESA Proba V satellite (the first operation in open space) launched in 2013 \cite{PROBAV}.

	Our proposal is to attach one or a few Timepix detectors outside  the shielding of the ''Gamma-400'' apparatus. The initial elliptic orbit of the apparatus will have the following parameters:  the apogee of about 300 000 km, the perigee of about 500 km, the inclination 51.8$^o$ and period of about 7 hours. Since the perigee lies low enough the apparatus will periodically cross both external and internal radiation belts. The Timepix detector could be able to monitor the flux of protons and electrons separately. The measurement of the energy spectrum could be available for lowenergetic component dominating in the outer part of the radiation belt. Assuming minimal frame length to be about 10 $\mu$s and typical cluster size of about 10 pixels we can estimate that the Timepix detector will be able to identify effectively protons and electrons at least up to the flux magnitude $7\times10^7$ cm$^{-1}$ s$^{-1}$ that corresponds to the occupancy of the detector on the level of 5\%. 
	
	In parallel with monitoring of the radiation belt the Timepix detector could be a unique instrument for investigation of charged particles flux in the outer region the Earth's magnetosphere. Such data can be combined with an information about solar activity and an information from the magnetometer planned to be installed at ''Gamma-400''. Study of energetic component of the solar wind out of the magnetosphere and its interaction with the magnetic field of our planet could be another promising task for the Timepix detector. This program will be especially important when the orbit will become more circular with a radius of about 100 000 km.

\end{document}